\documentclass[12pt,twoside]{article}
\usepackage{fleqn,espcrc1,epsfig}

\usepackage{graphicx}
\usepackage[figuresright]{rotating}

\newcommand{\AmS}{{\protect\the\textfont2
  A\kern-.1667em\lower.5ex\hbox{M}\kern-.125emS}}

\title{Traversing the QCD Phase Transition: 
Quenching Out of Equilibrium vs. Slowing Out of Equilibrium
vs. Bubbling Out of Equilibrium}

\author{Krishna Rajagopal\address{Center for Theoretical Physics,
MIT, Cambridge, MA 02139, USA}\thanks{Many thanks to
the organizers for a stimulating
conference in Adelaide. I am also
grateful to the Department of Energy's
Institute for Nuclear Theory at the University
of Washington for generous hospitality and support during
the writing of this contribution.  Research
supported in part by
a DOE OJI Award, by the A. P. Sloan Foundation and by the DOE
under agreement DE-FC02-94ER40818. Preprint MIT-CTP-2979.}}

\begin{document}
\maketitle

\begin{abstract}
I review arguments for the existence of a critical point $E$
in the QCD phase diagram as a function of temperature $T$ and
baryon chemical potential $\mu$. I describe how heavy ion collision
experiments at the SPS and RHIC can discover the tell-tale
signatures of such a critical point, thus mapping this
region of the QCD phase diagram.  I contrast the different
ways in which the matter produced in a heavy ion collision
can be driven out of equilibrium: quenching out of equilibrium
(possible, but not guaranteed, 
if the transition region is traversed at $\mu\ll\mu_E$) vs. 
slowing out of equilibrium (guaranteed for $\mu\sim\mu_E$) vs. bubbling
out of equilibrium (possible, but not guaranteed, for $\mu\gg\mu_E$). 
Quenching or bubbling create and amplify
distinct, detectable, non-gaussian fluctuations.
In contrast, slowing out of equilibrium reduces the magnitude
of the specific, detectable, gaussian fluctuations which signal
the presence of the critical point.  
\end{abstract}

\section{The Critical Point}

One goal of relativistic heavy ion collision experiments
is to explore and map the QCD phase diagram as a function
of temperature and baryon chemical potential.
Recent theoretical developments
suggest that a key qualitative feature, namely a critical
point which in a sense defines the landscape
to be mapped, may be within reach of discovery and analysis
by the CERN SPS or by RHIC, as data is taken at several different
energies \cite{SRS1,SRS2}.
The discovery of the critical point
would in a stroke transform the map of the QCD phase
diagram from one based only on
reasonable inference from universality, lattice gauge theory
and models into one with a solid experimental basis \cite{Review}.

In QCD with
two massless quarks ($m_{u,d}=0$; $m_s=\infty$)
the phase transition at which chiral symmetry is restored 
as $T$ is increased with $\mu=0$
is likely second order and belongs to the universality
class of $O(4)$ spin models in three dimensions \cite{piswil}.
Below $T_c$, chiral symmetry is broken and there are three
massless pions.  At $T=T_c$, there are four massless degrees
of freedom: the pions and the sigma. Above $T=T_c$, the pion
and sigma correlation lengths are degenerate and finite.

In nature, the light quarks are not massless.  Because
of this explicit chiral symmetry breaking,
the second order phase transition is replaced by an 
analytical crossover: physics changes dramatically but smoothly in the 
crossover region, and no correlation length diverges.
This picture is consistent with present lattice 
simulations \cite{latticereview},
which suggest $T_c\sim 140-190$ MeV \cite{latticeTc}.

\begin{figure}[t]
\vspace{-0.2in}
\begin{center}
\epsfig{file=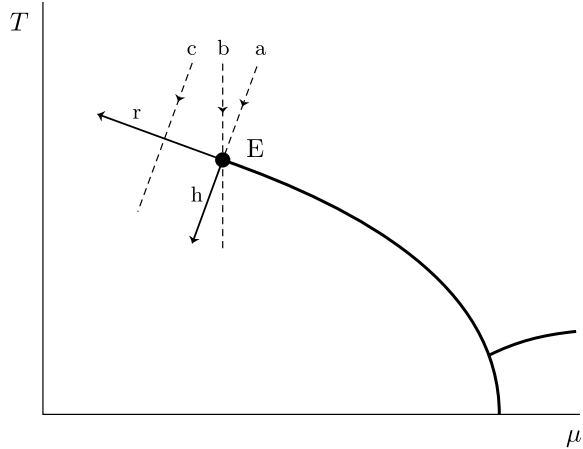,width=3.0in}
\end{center}
\vspace{-0.5in}
\caption{Sketch of the QCD phase diagram as a function of 
temperature $T$ and baryon chemical potential $\mu$.   
Chiral symmetry
is broken at low $T$ and $\mu$.  As $T$ is increased, chiral
symmetry is approximately restored via a smooth crossover to the
left of $E$ or a first order phase transition to the right of $E$.
The symmetry is only approximately restored because the light quarks
are not massless.
At the critical point $E$ at which the line
of first order phase transitions ends, the transition is second order 
and is in the Ising universality class.  
(At large $\mu$
and small $T$, there are color superconducting phases.)
The Ising model $r$-axis and $h$-axis and the trajectories a, b and c
will be discussed below.
}
\vspace{-0.2in}
\label{fig:phasediagram}
\end{figure}

Arguments based on a variety of 
models \cite{NJL,steph,ARW1,RappETC,bergesraj,stephetal,PisarskiRischke1OPT,CarterDiakonov}
indicate that the transition as a function of $T$ is first order at 
large $\mu$.\footnote{The realization that this phase transition
is not simply a transition at which the chiral condensate vanishes,
but is instead a transition at which a phase with nonzero chiral
condensate and a phase with nonzero superconducting condensate
compete and coexist has further strengthened the arguments in
favor of a first order phase 
transition on the 
$\mu$-axis in two-flavor 
QCD \cite{ARW1,bergesraj,PisarskiRischke1OPT,CarterDiakonov}.}   
This suggests that the
phase diagram features a critical point $E$ at which
the line of first order phase transitions present for 
$\mu>\mu_E$ ends, as shown in Figure 1.\footnote{If
the up and down quarks were massless, $E$ would
be a tricritical point, at which the first
order transition becomes second order.}
At $\mu_E$, the phase transition at $T=T_E$ is second order
and is in the Ising universality class \cite{bergesraj,stephetal}.
Although the
pions remain massive, the correlation length in the $\sigma$ channel
diverges due to universal long wavelength fluctuations
of the order parameter.
This results in characteristic signatures,
analogues of critical opalescence in the sense that they
are unique to collisions which freeze out near the
critical point, which
can be used to discover $E$ \cite{SRS1,SRS2}.

The position of the critical point is, of course, not universal.
Furthermore, it is sensitive to the value of the strange
quark mass.  $\mu_E$ decreases as $m_s$ is decreased \cite{SRS1},
and at some $m_s^c$, it reaches $\mu_E=0$ and the transition
becomes entirely first order \cite{rajwil}.  
The value of $m_s^c$ is an open question,
but lattice simulations suggest that it is about half the
physical strange quark mass \cite{columbia,kanaya}, although 
these results are not yet conclusive \cite{oldkanaya}.
Of course, experimentalists cannot vary $m_s$.  They
can, however, vary $\mu$.  The AGS, with beam energy 11 AGeV
corresponding to
$\sqrt{s}=5$ GeV, creates fireballs which freeze out
near $\mu\sim 500-600$ MeV \cite{PBM}.  
When the SPS runs with $\sqrt{s}=17$ GeV
(beam energy 158 AGeV),
it creates fireballs which freeze out near $\mu\sim 200$ MeV \cite{PBM}.
RHIC will make even smaller values of $\mu$ accessible.
By dialing $\sqrt{s}$ and thus $\mu$, experimenters can find
the critical point $E$.  Finding $E$ would confirm that
$m_s^c$ is less than the physical value of $m_s$, and 
that the phase transition at lower chemical potentials ($\mu<\mu_E$)
is a smooth crossover.

\section{Discovering the Critical Point}

Predicting
$\mu_E$, and thus suggesting the $\sqrt{s}$ to
use to find $E$, is beyond the reach of present
theoretical methods
because
$\mu_E$ is both nonuniversal and sensitively dependent on the mass
of the strange quark.   
Crude models suggest that $\mu_E$ could be $\sim 600-800$ MeV
in the absence of the strange quark \cite{bergesraj,stephetal}; 
this in turn suggests that
in nature $\mu_E$ may have of order half this value, and may therefore
be accessible at the SPS if the SPS 
runs with $\sqrt{s}<17$ GeV.   However, at present theorists cannot
predict the value of $\mu_E$ even to within a factor of two.
The SPS can search a significant fraction of the parameter
space; if it does not find $E$, it will then be up to 
the RHIC experiments to map the $\mu_E< 200 $ MeV region.

Locating $E$ on the phase diagram can only be done convincingly by an
experimental discovery.  Theorists can, however, do reasonably well at
describing the phenomena that occur near $E$, thus enabling
experimenters to locate it.  This is the goal of Ref.  \cite{SRS2}.
The signatures proposed there are based on the fact that $E$ is a
genuine thermodynamic singularity at which susceptibilities diverge
and the order parameter fluctuates on long wavelengths. The resulting
signatures are {\it nonmonotonic} as a function of $\sqrt{s}$: as this
control parameter is varied, we should see the signatures strengthen
and then weaken again as the critical point is approached and then
passed.

The simplest observables to use are the event-by-event fluctuations of
the mean transverse momentum of the charged particles in an event,
$p_T$, and of the total charged multiplicity in an event, $N$.  
The fluctuations observed in SPS collisions by NA49
are as perfect gaussians as the data statistics
allow, as expected for freeze-out from a system in thermal 
equilibrium \cite{NA49}.
The data on multiplicity fluctuations show evidence for
a nonthermodynamic contribution, which is to 
be expected since the extensive quantity $N$ is sensitive
to the initial size of the system and thus to nonthermodynamic
effects like variation in impact parameter. The
contribution of such effects to the fluctuations
have now been estimated \cite{BH,DS}; the combined thermodynamic
and nonthermodynamic fluctuations are in satisfactory agreement
with the data \cite{DS}.
The width 
of the event-by-event distribution 
of the mean $p_T$ of the charged pions in a single 
event\footnote{This width can be measured even
if one observes only two pions per event \cite{bialaskoch};
large acceptance data as from NA49 is required in order
to learn that the distribution is gaussian, that 
thermodynamic predictions may be valid, and that
the width is therefore the only interesting quantity to measure.}
is in good agreement with predictions based on noncritical thermodynamic 
fluctuations \cite{SRS2}.
That is, NA49 data are consistent with the hypothesis
that almost all the observed event-by-event fluctuation in 
mean $p_T$, an intensive quantity,
is thermodynamic in origin.
This bodes well for the detectability of systematic
changes in thermodynamic fluctuations near $E$.

One 
analysis described in detail in Ref. \cite{SRS2} is based on the ratio
of the width of the true event-by-event distribution of the mean $p_T$
to the width of the distribution in a sample of mixed events. This
ratio was called $\sqrt{F}$. NA49 has measured $\sqrt{F}=1.002\pm
0.002$ \cite{NA49,SRS2}, which is consistent with expectations for
noncritical thermodynamic fluctuations.\footnote{In an infinite system
made of classical particles which is in thermal equilibrium,
$\sqrt{F}=1$.  Bose effects increase $\sqrt{F}$ by $1-2\%$
\cite{Mrow,SRS2}; an anticorrelation introduced by energy conservation
in a finite system --- when one mode fluctuates up it is more likely
for other modes to fluctuate down --- decreases $\sqrt{F}$ by $1-2\%$
\cite{SRS2}; two-track resolution also decreases $\sqrt{F}$ by $1-2\%$
\cite{NA49}. The contributions due to correlations introduced by
resonance decays and due to fluctuations in the flow velocity are each
much smaller than $1\%$ \cite{SRS2}.}  Critical fluctuations
of the $\sigma$ field, i.e. the characteristic long wavelength
fluctuations of the order parameter near $E$, influence pion momenta
via the (large) $\sigma\pi\pi$ coupling and increase $\sqrt{F}$
\cite{SRS2}.  The effect is proportional to $\xi_{\rm freezeout}^2$,
where $\xi_{\rm freezeout}$ is the $\sigma$-field correlation length
of the long-wavelength fluctuations at freezeout \cite{SRS2}.  If
$\xi_{\rm freezeout}\sim 3$ fm, the ratio $\sqrt{F}$ increases by
$\sim 3-5\%$, ten to twenty times the statistical error in the present
measurement \cite{SRS2}.  This observable is valuable because data on
it has been analyzed and presented by NA49, and it can therefore be
used to learn that Pb+Pb collisions at 158 AGeV do {\it not} freeze
out near~$E$.

Once $E$ is located, however, other observables which 
are more sensitive to critical effects will be more useful.
For example, a $\sqrt{F_{\rm soft}}$,
defined using only the softest $10\%$ of the pions in each event, 
will be much more sensitive to the critical long wavelength 
fluctuations.  The higher $p_T$ pions are less affected
by the $\sigma$ fluctuations \cite{SRS2}, 
and these relatively unaffected pions
dominate the mean $p_T$ of all the pions in the
event.  This is why the increase in $\sqrt{F}$ near the critical point 
will be much less than that of $\sqrt{F_{\rm soft}}$. 

The multiplicity of soft pions is an 
example of an observable which may
be used to detect the critical fluctuations 
without an event-by-event analysis.
The post-freezeout decay of sigma mesons, which are copious
and light at freezeout near $E$ and which
decay subsequently when their mass increases above
twice the pion mass, should result in a population of pions 
with $p_T\sim m_\pi/2$ which appears only for freezeout
near the critical point \cite{SRS2}.  
If $\xi_{\rm freezeout}> 1/m_\pi$, this population
of unusually low momentum pions will be comparable in
number to that of the ``direct'' pions (i.e. those which
were pions at freezeout) and will result in a large
signature. 

The variety of observables
which should {\it all} vary nonmonotonically with $\sqrt{s}$
(and should all peak at the same $\sqrt{s}$)
is sufficiently great that if it were to turn out that 
$\mu_E<200$ MeV, making $E$ inaccessible to the SPS, all four
RHIC experiments could play a role in the study of the critical
point.

\section{Slowing Out of Equilibrium: How Large Can $\xi$ Grow?  }

The purpose of Ref. \cite{Berdnikov} 
is to estimate how large 
$\xi_{\rm freezeout}$ can become, thus making the predictions
of Ref. \cite{SRS2} for the magnitude of various signatures
more quantitative.  In an ideal system of infinite size
which was held at $T=T_E$; $\mu=\mu_E$ for an infinite time,
the correlation length $\xi$ would be infinite.  Ref. \cite{SRS2}
estimated that finite size effects limit $\xi$ to be 
about 6 fm at most. We argue in Ref. \cite{Berdnikov} that
limitations imposed by the finite duration of a heavy
ion collision are more severe, preventing
$\xi$  from growing larger than about $2/T_E \sim 3$ fm.

The nonequilibrium dynamics analyzed in Ref. \cite{Berdnikov}
is {\it guaranteed}
to occur in a heavy ion collision which passes near $E$,
even if local thermal equilibrium is achieved 
at a higher temperature during the earlier evolution
of the plasma created in the collision.  
We assume early thermal (although not necessarily chemical)
equilibration, and ask how the system evolves
out of equilibrium as it passes $E$.  
For the present, assume that the system cools 
through the critical point $E$ as sketched in trajectory (b)
of Figure 1.
If it were to cool arbitrarily slowly, $\xi(T)=\xi_{\rm eq}(T)$
would be maintained at all temperatures, and $\xi$ would
diverge at $T_E$.  However, it would take an infinite
time for $\xi$ to grow infinitely large.  Indeed, near
a critical point, the longer the correlation length, the
longer the equilibration time, and the slower the 
correlation length can grow. This critical slowing
down means that the
correlation length cannot grow as fast as $\xi_{\rm eq}$,
and the system cannot stay in equilibrium.  

We use the theory of dynamical critical phenomena to 
describe the effects of critical slowing down
on the time development of the correlation length $\xi(t)$,
using the three-dimensional Ising model to describe 
$\xi_{\rm eq}(T,\mu)$ near $E$ \cite{Berdnikov}.
In the Ising model, the order parameter and the correlation
length are functions of the reduced temperature 
$r$ and the magnetic field $h$.  (In the Ising model,
$r$ is defined as $(T-T_c)/T_c$ and is
usually called $t$;
we reserve the symbol $t$ for time, however.)  We assume that 
the $r$- and $h$-axes are as sketched in Figure 1, although in reality
they will be deformed.  We consider cooling along trajectories 
that pass through $E$ at various angles,
like (a) and (b) sketched in Figure 1 and also consider trajectories
like (c) which come near to but miss $E$. 
Our results depend on the universal 
function describing $\xi_{\rm eq}(r,h)$, on the universal dynamical exponent
$z$ describing critical slowing down (perturbations away from
equilibrium relax toward equilibrium on a timescale which
scales with $\xi$ like $A\xi^z$ \cite{HoHa}), on the nonuniversal
constant $A$, the nonuniversal constants  
which relate $r$ and $h$ to $(T-T_E)$ and $(\mu-\mu_E)$, on $T_E$
which we take to be $\sim 140$ MeV,
and finally on the cooling rate $|dT/dt|$ which
we estimate to be 4 MeV/fm \cite{HS,Heinz,Bravina,Berdnikov}.

Our results 
have a number of consequences, detailed in the remainder
of this section,  which should
be taken into account both in planning experimental
searches for the QCD critical point, and in planning future
theoretical work.

We do indeed find that because of the critical slowing down of 
the long wavelength
dynamics near $E$, the correlation length does not have time
to grow as large as it would in equilibrium:  
we find 
$\xi_{\rm freezeout}\sim 2/T_E \sim 3$ fm for trajectories
passing near $E$.

Although critical slowing down hinders the growth of $\xi$, it 
also slows the decrease of $\xi$  as the system continues to cool 
below the critical point.  As 
a result, $\xi$ does not
decrease significantly between the phase transition and 
freezeout.

Our estimate that $\xi$ does not
grow larger than $2/T_E$ is robust in three senses.  First, it depends
very little on the angle with which the trajectory passes
through $E$. Second, 
it turns out to depend on only one combination of all the
nonuniversal quantities which play a role. We call this
parameter $a$; it is proportional to $|dT/dt|^{-1}$. 
Third, our results do not depend sensitively on $a$. 
We show that the maximum value of $\xi$ scales like $a^{\frac{\nu/\beta\delta}
{1+z\nu/\beta\delta}}\approx a^{0.215}$ 
\cite{Berdnikov}.\footnote{A scaling law
of this form (but of course with different exponents) 
relating the maximum correlation length which is
reached to the cooling rate with which a second order
phase transition is traversed 
was first discovered in the theory of defect formation at a 
second order phase transition \cite{Zurek}. 
In this context,
the maximum correlation length reached during the phase transition
sets the scale for the initial separation between defects --- vortices
in a superfluid or in liquid crystals, for example ---
created as the system cools through the 
transition. 
This initial network of defects coarsens at later times.
The Ising phase transition of interest to us 
creates no defects. It is nevertheless
very pleasing that scaling laws analogous to the one we
need have been tested quantitatively in 
numerical simulations of defect formation and 
dynamics \cite{Antunes} and, furthermore, are supported by 
data from experiments on
liquid crystals \cite{LiquidCrystal} and superfluid $^3$He \cite{Helium}.}
Thus, for example,
$|dT/dt|$ would have to be a factor of 25 smaller than 
we estimate in order
for $\xi$ to grow to $4/T_E$ instead of $2/T_E$.
Although our results are robust in this sense,
they cannot be treated as precise because 
our assumption that the dynamics of $\xi$ in QCD 
is described by the universal classical dynamics
of the three-dimensional Ising model only becomes precise if $\xi\gg 1/T_E$,
while our central result is that $\xi$ does not grow beyond $\sim 2/T_E$.
A $3+1$-dimensional quantum field theoretical treatment of 
the interplay between cooling and 
the dynamics of critical slowing down is not
yet available, but promising first steps in this direction 
can be found in Ref. \cite{Boyan}.

A result which is of great importance in the planning
of experimental searches 
is that one need not hit $E$ precisely in order
to find it.  Our analysis of
trajectories like (c) of Figure 1 demonstrates that if one were to
do a scan with collisions at many finely spaced values of
the energy and thus $\mu$, one would see signatures of $E$
with approximately the same magnitude over a broad range
of $\mu$.  The magnitude of the 
signatures will not be narrowly peaked as $\mu$ is varied.
As long as one gets close enough to $E$ that the equilibrium
correlation length is $(2-3)/T_E$, the actual correlation
length $\xi$ will grow to $\sim 2/T_E$.  There is no advantage
to getting closer to $E$, because critical slowing down
prevents $\xi$ from getting much larger even if $\xi_{\rm eq}$ does.
Data at many finely spaced values of $\mu$ is
{\it not} called 
for.\footnote{Analysis within the toy model of Ref. \cite{bergesraj}
suggests that in the absence of the strange quark, the
range of $\mu$ over which $\xi_{\rm eq}> 2$ fm is about
$\Delta \mu \sim 120$ MeV for $\mu_E \sim 800$ MeV.
Similar results can be obtained \cite{Misha} within 
a random matrix model \cite{stephetal}.  It is likely
over-optimistic to estimate $\Delta \mu \sim 120$ MeV
when the effects of the strange quark are included
and $\mu_E$ itself is reduced. A conservative estimate would
be to use the models to estimate that $\Delta \mu/\mu_E \sim 15\%$.}

Knowing that we are
looking for  $\xi_{\rm freezeout} \approx 3$ fm is very helpful
in suggesting 
how to employ the signatures described in detail
in Ref. \cite{SRS2}.    The excess of pions with $p_T\sim m_\pi/2$
arising from post-freezeout decay of sigmas is large 
as long as $\xi_{\rm freezeout}\sim 1/m_\pi$, and does not
increase much further if $\xi_{\rm freezeout}$ is longer.
This makes it an ideal signature.
The increase in the
event-by-event fluctuations in the mean transverse
momentum of the charged pions in an event (described
by the ratio $\sqrt {F}$ of Ref. \cite{SRS2}) is proportional
to $\xi_{\rm freezeout}^2$.  The results of Ref. \cite{SRS2} suggest that
for $\xi_{\rm freezeout}\sim 3$ fm, this will be a $3-5\%$ effect.
This is ten to twenty times larger than the statistical error in the
present NA49 data, but not so large as to make one confident
of using this alone as a signature for $E$.   The solution
is to use signatures which focus on the event-by-event
fluctuations of only the low momentum pions. Unusual
event-by-event
fluctuations in the pion momenta arise via the coupling
between the pions and the sigma order parameter which, at freezeout, is
fluctuating with correlation length $\xi_{\rm freezeout}$.
This interaction has the largest effect on the softest 
pions \cite{SRS2}.
$\sqrt{F_{\rm soft}}$,
described in the previous section, is a good example of an observable
which takes advantage of this.  Depending on the details of the
cuts used to define it, it should be enhanced by many tens
of percent in collisions passing near $E$.
Ref. \cite{SRS2} suggests other such observables,
and more can surely be found.  Together, the excess
multiplicity at low momentum (due to post-freezeout sigma decays) and
the excess event-by-event fluctuation of the momenta of
the low momentum pions (due to their coupling to the order parameter
which is fluctuating with correlation length $\xi_{\rm freezeout}$)
should allow a convincing detection of the critical point $E$.
Both should behave nonmonotonically as the collision energy,
and hence $\mu$, are varied.  Both should peak for those
heavy ion collisions which freeze out near $E$, with
$\xi_{\rm freezeout}\sim 3$ fm.

\section{Quenching Out of Equilibrium vs. Slowing Out of Equilibrium
vs. Bubbling Out of Equilibrium}

As the matter created in a heavy ion collision
cools  through the 
QCD phase transition, it may be driven out of equilibrium.
The way in which this may happen is quite different depending
on whether the phase transition is traversed with 
$\mu\gg\mu_E$, $\mu\ll\mu_E$ or $\mu\sim\mu_E$.

If $\mu>\mu_E$, the phase transition is first order; if 
$(\mu-\mu_E)$ is large enough, the transition may be sufficiently
strongly first order that it proceeds via the nucleation of 
well-separated bubbles of hadronic matter, which then grow.
This bubbling results in a spatially inhomogeneous
hadronic system.  If the system freezes out before it gets
re-homogenized, the result will be large, non-gaussian fluctuations
in hadron multiplicity as a function of rapidity \cite{Bubbling}.
These overdense and underdense regions in rapidity would provide
a distinctive signature, and have not been seen.

If $\mu\ll\mu_E$, the phase transition is a smooth crossover. 
This is the region of the diagram where the transition
may be traversed most quickly, with the most rapid cooling 
rate.
(For $\mu\sim\mu_E$, the specific heat is unusually large and therefore
even though the rate of decrease of energy density is not unusual, the 
cooling rate $|dT/dt|$
is unusually small. For $\mu\gg\mu_E$,
neither of the two phases which coexist at the first order phase
transition has an unusually large specific heat. The system
nevertheless spends a long time at the coexistence
temperature $T_c$ because of the release of latent heat.  Thus,
$|dT/dt|$ is allowed to be largest for $\mu\ll\mu_E$.)  
Although the smooth crossover with nondivergent specific heat
allows a large $|dT/dt|$ in principle, the cooling rate
which is actually achieved will depend on the system size and 
on the duration of time between the initial collision and the
phase transition.
If conditions can be found in which $|dT/dt|$
can be made large enough, the dynamics may 
then be well-modelled as a quench,
in which most of the degrees of freedom in the system
are imagined to cool arbitrarily rapidly.  In this circumstance,
the long wavelength modes of the pion field become unstable
to exponential growth \cite{quench}, leading to the formation
of long wavelength pion oscillations.  Once excited, these
oscillations may be further amplified by parametric 
resonance \cite{parametricresonance}.\footnote{As seems 
indicated \cite{parametricresonance}
by the long period of time over which amplification is
seen in the simulations of Ref. \cite{quench}.}
These long wavelength
disorientations of the chiral condensate result in characteristic
fluctuations in the ratio of the number of low $p_T$ neutral 
pions to the number of low $p_T$ charged pions \cite{DCC,quench,DCCreviews}.
Other signatures of DCC oscillations
have also been discussed \cite{othersignatures}.
WA98 \cite{WA98} and NA49 \cite{NA49} have looked for
signatures of DCC oscillations in 158 AGeV PbPb
collisions, but none have been 
seen.

The considerations
above suggest that RHIC collisions provide 
more favorable conditions for DCC production
than SPS collisions, because $\mu$ is smaller and one is more
likely to have a smooth crossover which
allows large $|dT/dt|$ .  However, the hadronic systems
produced in RHIC collisions will be large, will traverse the phase
transition late, and will freeze out 
even later. The late transition reduces the cooling rate, making DCC formation
less likely, and the late freeze out makes wash-out 
of DCC signatures a worry.  This
suggests that the most favorable conditions in which to 
look for DCC signatures may be RHIC collisions of medium-sized ions,
or AuAu collisions at RHIC with impact parameters of order the
nuclear radius.  (That is, midway between central and peripheral collisions.)
Quite different considerations have lead Asakawa, Minakata and M\"uller
to propose that non-central RHIC collisions provide 
the most favorable conditions 
for DCC formation \cite{anomalykick}: 
these collisions
feature strong 
electromagnetic $\bf E \cdot \bf B$ fields which may, 
through the anomaly,  result in a coherent excitation of the $\pi^0$
field.  There are therefore three reasons to suggest 
that non-central RHIC collisions may yield DCC signatures:
the smooth crossover and small system size 
allow the system to cool through the transition faster, favoring 
the instability driven growth of DCCs; the electromagnetic
kick may further drive their growth; and, the smaller system size
will hasten freezeout.  The first and third reasons also apply
to RHIC collisions of medium-sized ions.

For $\mu\sim\mu_E$, the specific heat is larger in the vicinity
of the critical point than anywhere else on the phase diagram,
and the cooling rate $|dT/dt|$ is correspondingly slow.
This slow cooling means that those collisions which create matter
which cools through the transition region with $\mu\sim\mu_E$
are {\it least} likely to be driven far from
equilibrium.  That is, we do not expect DCC signatures or signatures
of bubbling in these collisions.  The interesting twist
described in the previous section
is that it is nevertheless true that such collisions cannot
stay {\it precisely} in equilibrium \cite{Berdnikov}.  
Even though the system cools through the vicinity
of $E$ slowly, the correlation length still grows much
too slowly to become as long
as it would in equilibrium.  We have discussed 
this phenomenon of slowing out of equilibrium
at length, as it determines the magnitude of
signatures of the critical point. 
Here, we note that this non-equilibrium physics is
much milder than quenching or bubbling.  Quenching or bubbling 
create {\it more} fluctuations than would be obtained in
thermal equilibrium.  They also create fluctuations which
are non-gaussian in characteristic ways.  In contrast, 
slowing out of equilibrium serves to {\it reduce} the growth
of fluctuations relative to what would be achieved if 
equilibrium were maintained.  
The critical point $E$ will be signalled by an increase in
the magnitude of suitably chosen event-by-event fluctuations,
for example parametrized by the growth in $\sqrt{F}$ 
and $\sqrt{F_{\rm soft}}$,
but these fluctuations are expected to remain gaussian.
Slowing out of equilibrium means that these gaussian
fluctuations are smaller than they would be if 
equilibrium were maintained near $E$. They are nevertheless
larger than the equilibrium thermal fluctuations far from $E$
and hence result in nonmonotonic signatures which peak for collisions which
freezeout near $E$.\footnote{In addition to 
restricting the growth of fluctuations, critical slowing down retards
the decrease of the fluctuations after the transition.
In addition, the large specific heat and consequent slow cooling rate
result in an elevated freezeout temperature \cite{SRS2}.  This suggests
that signatures of $E$ will not be washed out before freezeout;
if necessary, however, smaller ions or non-central collisions
can be used to further elevate the freezeout 
temperature \cite{laterfreezeout,SRS2}.}

We have learned much from the beautiful gaussian event-by-event
fluctuations observed by NA49.  The magnitude of these flucutations
are consistent with the hypothesis that the hadronic system
at freezeout is in approximate thermal equilibrium. These and
other data show no signs that the system at freezeout has
recently been bubbled or quenched.  There is also no sign of
the larger, but still gaussian, fluctuations which would signal
freezeout near the critical point $E$.  Combining these 
observations with the observation of several tantalizing indications
that the matter created in SPS collisions is not well described
at early times by hadronic models \cite{HeinzJacob} suggests
that collisions at the SPS may be exploring the crossover region
to the left of the critical point $E$, in which
the matter is not well-described as a hadron gas but
is also not well-described as a quark-gluon plasma.  This speculation
could be confirmed in two ways.   First, if the SPS is probing
the crossover region then the coming experiments
at RHIC may discover direct signatures of an early partonic phase, 
which are well-described by theoretical calculations beginning from 
an equilibrated quark-gluon plasma.  Second, 
if 158 AGeV collisions
are probing the
crossover region not far to the left of the critical point $E$, 
then running at lower energies
would result in the discovery of $E$. 
If, instead, RHIC were to discover
$E$ with $\mu_E<200$~MeV, that would indicate that the SPS 
experiments have probed the weakly first order region 
just to the right of $E$. Regardless, 
discovering $E$ would take all the speculation
out of mapping this part of the QCD phase diagram.


\begin{thebibliography}{9}

\bibitem{SRS1}
M.~Stephanov, K.~Rajagopal and E.~Shuryak,
Phys.\ Rev.\ Lett.\  {\bf 81}, 4816 (1998).

\bibitem{SRS2}
M.~Stephanov, K.~Rajagopal and E.~Shuryak,
Phys.\ Rev.\  {\bf D60}, 114028 (1999).


\bibitem{Review} 
For a review, see K. Rajagopal, 
Nucl. Phys. {\bf A661}, 150 (1999) [hep-ph/9908360].

\bibitem{piswil}
R. Pisarski and F. Wilczek, Phys. Rev. {\bf D29}, 338 (1984).

\bibitem{latticereview}
For reviews, see 
F.~Karsch,
hep-lat/9909006;
E. Laermann Nucl. Phys. Proc. Suppl. {\bf 63}, 114 (1998); 
A. Ukawa, Nucl. Phys. Proc. Suppl. {\bf 53}, 106 
(1997).

\bibitem{latticeTc} For example, S. Gottlieb et al., 
Phys. Rev. {\bf D55}, 6852 (1997); R.~Mawhinney, talk
at ISMD99, Providence, RI, 1999; F.~Karsch,
hep-lat/9909006.

\bibitem{NJL}  
A. Barducci,
R. Casalbuoni, S. DeCurtis, R. Gatto, G. Pettini, Phys. Lett. {\bf B231},
463 (1989);
S.P. Klevansky, Rev. Mod. Phys. {\bf 64}, 649 (1992); 
A. Barducci, R.~Casalbuoni, G. Pettini and R. Gatto, Phys. Rev. {\bf D49},
426 (1994).

\bibitem{steph} 
        M. Stephanov, Phys. Rev. Lett. {\bf 76}, 4472 (1996);
        Nucl. Phys. Proc. Suppl. {\bf 53}, 469 (1997).

\bibitem{ARW1}
M.~Alford, K.~Rajagopal and F.~Wilczek,
Phys.\ Lett.\  {\bf B422}, 247 (1998).

\bibitem{RappETC}
R. Rapp, T. Sch\"afer,
E. V. Shuryak and M. Velkovsky, Phys. Rev. Lett. {\bf 81}, 53 (1998).

\bibitem{bergesraj} 
J.~Berges and K.~Rajagopal,
Nucl.\ Phys.\  {\bf B538}, 215 (1999).

\bibitem{stephetal}
M. A. Halasz, A. D. Jackson, R. E. Shrock, M. A. Stephanov
and J.~J.~M.~Verbaarschot, Phys. Rev. {\bf D58}, 096007 (1998).

\bibitem{PisarskiRischke1OPT}
R. Pisarski and D. Rischke, Phys. Rev. Lett. {\bf 83}, 37
(1999).

\bibitem{CarterDiakonov}
G. Carter and D. Diakonov, Phys. Rev. {\bf D60}, 016004 (1999).

\bibitem{rajwil}
F. Wilczek, Int. J. Mod. Phys. {\bf A7}, 3911 (1992);
K. Rajagopal and F.~Wilczek, Nucl. Phys. {\bf B399}, 395 (1993).

\bibitem{columbia}
F. Brown {\it et al}, Phys. Rev. Lett. {\bf 65}, 2491 (1990).

\bibitem{kanaya}
JLQCD Collaboration, Nucl. Phys. Proc. Suppl. {\bf 73}, 459 (1999).

\bibitem{oldkanaya}
Y. Iwasaki {\it et al}, Phys. Rev. {\bf D54}, 7010 (1996).

\bibitem{PBM}
See, e.g., P. Braun-Munziger and J. Stachel,
Nucl. Phys. {\bf A606}, 320 (1996).

\bibitem{NA49}
H.~Appelshauser {\it et al.}  [NA49 Collaboration],
Phys.\ Lett.\  {\bf B459}, 679 (1999).

\bibitem{BH}
G.~Baym and H.~Heiselberg,
Phys.\ Lett.\  {\bf B469}, 7 (1999).

\bibitem{DS}
G. Danilov and E. Shuryak, nucl-th/9908027.

\bibitem{bialaskoch}
A. Bialas and V. Koch, Phys. Lett. {\bf B456}, 1 (1999).

\bibitem{Mrow} 
St. Mr\'owczy\'nski, Phys. Lett. {\bf B430}, 9  (1998).

\bibitem{Berdnikov}
B.~Berdnikov and K.~Rajagopal,
hep-ph/9912274.

\bibitem{HoHa} 
P. C. Hohenberg and B. I. Halperin, Rev. Mod. Phys. {\bf 49}, 435 (1977).

\bibitem{HS} C. M. Hung and E. Shuryak, Phys. Rev. {\bf C57}, 1891 (1998).

\bibitem{Heinz}
E. Schnedermann and U. Heinz, Phys. Rev. {\bf C47}, 1738 (1993); and 
{\bf C50}, 1675 (1994); 
B. Tomasik, U. A. Wiedemann and U. Heinz, nucl-th/9907096; 
U. Heinz, private communication.

\bibitem{Bravina}
L. V. Bravina {\it et al},  Phys. Rev. {\bf C60}, 024904 (1999).

\bibitem{Zurek}
W. H. Zurek, Nature {\bf 317}, 505 (1985); Phys. Rept. {\bf 276}, 177 (1996).

\bibitem{Antunes}
N.~D.~Antunes, L.~M.~A.~Bettencourt and W.~H.~Zurek,
Phys.\ Rev.\ Lett.\  {\bf 82}, 2824 (1999), and references therein.

\bibitem{LiquidCrystal}
I. Chuang, R. Durrer, N. Turok and B. Yurke, Science {\bf 251}, 1336 (1991);
M. J. Bowick, L. Chander, E. A. Schiff and A. M. Srivastava, {\it ibid.}
{\bf 263}, 943 (1994).

\bibitem{Helium}
C. B\"auerle {\it et al.}, Nature {\bf 382}, 332 (1996);
V. M. H. Ruutu {\it et al.}, Nature {\bf 382}, 334 (1996); Phys. Rev. Lett.
{\bf 80}, 1465 (1998); V. B. Eltsov, M. Krusius and 
G. E. Volovik cond-mat/9809125.

\bibitem{Boyan}
D. Boyanovsky, H. J. de Vega and M. Simionato, hep-ph/0004159.

\bibitem{Misha}
M. Stephanov, private communication.

\bibitem{Bubbling}
See, for example, H. Heiselberg and A. D. Jackson,
nucl-th/9809013;
I.~N.~Mishustin,
Phys.\ Rev.\ Lett.\  {\bf 82}, 4779 (1999).

\bibitem{quench}
K.~Rajagopal and F.~Wilczek,
Nucl.\ Phys.\  {\bf B404}, 577 (1993).

\bibitem{parametricresonance}
S.~Mrowczynski and B.~M\"uller,
Phys.\ Lett.\  {\bf B363}, 1 (1995);
H.~Hiro-Oka and H.~Minakata,
Phys.\ Lett.\  {\bf B425}, 129 (1998);
D.~I.~Kaiser,
Phys.\ Rev.\  {\bf D59}, 117901 (1999);
H.~Hiro-Oka and H.~Minakata,
Phys.\ Rev.\  {\bf C61}, 044903 (2000).

\bibitem{DCC}
A.~A.~Anselm,
Phys.\ Lett.\  {\bf B217}, 169 (1989);
A.~A.~Anselm and M.~G.~Ryskin,
Phys.\ Lett.\  {\bf B266}, 482 (1991); J.-P. Blaizot and A. Krzywicki,
Phys. Rev. {\bf D46}, 246 (1992); J. D. Bjorken, Int.
J. Mod. Phys. {\bf A7}, 4189 (1992); J. D. Bjorken, Acta Phys. Pol. 
{\bf B23}, 561 (1992); K. L. Kowalski and C. C. Taylor, hep-ph/9211282.

\bibitem{DCCreviews}
For reviews, see K.~Rajagopal,
hep-ph/9504310;
J.~P.~Blaizot and A.~Krzywicki,
Acta Phys.\ Polon.\  {\bf 27}, 1687 (1996);
K.~Rajagopal,
hep-ph/9703258.

\bibitem{othersignatures}
C.~Greiner, C.~Gong and B.~M\"uller,
Phys.\ Lett.\  {\bf B316}, 226 (1993);
Z.~Huang, M.~Suzuki and X.~Wang,
Phys.\ Rev.\  {\bf D50}, 2277 (1994);
S.~Gavin,
Nucl. Phys. {\bf A590}, 163 (1995);
D.~Boyanovsky, H.~J.~de Vega, R.~Holman and S.~Prem Kumar,
Phys.\ Rev.\  {\bf D56}, 5233 (1997);
Y.~Kluger, V.~Koch, J.~Randrup and X.~Wang,
Phys.\ Rev.\  {\bf C57}, 280 (1998);
C.~Chow and T.~D.~Cohen,
Phys.\ Rev.\  {\bf C60}, 054902 (1999).

\bibitem{WA98}
M.~M.~Aggarwal {\it et al.}  [WA98 Collaboration],
Phys.\ Lett.\  {\bf B420}, 169 (1998).

\bibitem{anomalykick}
H.~Minakata and B.~M\"uller,
Phys.\ Lett.\  {\bf B377}, 135 (1996);
M.~Asakawa, H.~Minakata and B.~M\"uller,
Nucl.\ Phys.\  {\bf A638} (1998) 443C;
M.~Asakawa, H.~Minakata and B.~M\"uller,
Phys.\ Rev.\  {\bf D58}, 094011 (1998).

\bibitem{laterfreezeout} The earlier freeze out of smaller systems 
has been confirmed experimentally by observing the $A$-dependence
of the freeze-out temperature via analyses of flow \cite{HS}, Coulomb
effects (H. W. Barz, J.~P.~Bondorf, J. J. Gaardhoje and H. Heiselberg,
Phys. Rev. {\bf C57} (1998) 2536), 
and pion interferometry (U. Heinz, Nucl. Phys. {\bf A638}, 357c (1998).)

\bibitem{HeinzJacob}
Reviewed in U.~Heinz and M.~Jacob,
nucl-th/0002042.


\end{thebibliography}
\end{document}